\documentclass{article}

\input{tcilatex}

\begin{document}

Soviet Physics JETP; volume 30, number 3, pp. 512-513; \ March 1970

\bigskip

\textbf{NONLINEAR EVOLUTION OF DISTURBANCES IN A (1+1)-DIMENSIONAL UNIVERSE}

\bigskip

E. A. Novikov

\bigskip

Institute of Atmospheric Physics, USSR Academy of Sciences

Submitted April 4, 1969

Zh. Exper. Teor Fiz. \textbf{57}, 938-940 (September, 1969).

\bigskip

\bigskip

General exact solution is obtained for the problem of the development of
arbitrary disturbances of the density and velocity in a (1+1)-dimensional
universe. This analytical solution may serve, particularly, as a test for
numerical methods. For an illustration, the nonlinear distortion of a
sinusoidal perturbation of the initial density is calculated.

\bigskip

\bigskip

\textbf{1.} The problem of the origin of galaxies and their distributions
with respect to masses and angular momenta is, first of all, a problem of
the nonlinear evolution of disturbances of the density. The spherically
symmetric solution by R. Tolman [1] and the first-order perturbations to the
Friedmann solution, which were obtained by E. M. Lifshitz [2], are well
known. These results, and also the formulation of the problem about the
origin of galaxies, are significantly simplified in the Newtonian
approximation [3,4], and also the Newtonian approximation turns out to be
adequate for the problem under consideration, at least for the epoch after
recombination of the plasma. But even in the Newtonian approximation the
problem reduces to a rather complicated system of nonlinear equations with
partial derivatives [4], which must be investigated by approximate methods
(using different orders of perturbation theory and numerical procedures). In
this connection it is useful to have a simple model of the problem
possessing an exact solution, which may serve as a test of the approximate
methods.

\textbf{2.} Let us consider a (1+1)-dimensional universe filled with dust
(we shell be interested in scales of distance which are larger than the
Jeans wavelength). The basic equations (corresponding to a special case of
equations in Ref. 4) have the form:

\begin{equation}
\frac{\partial v}{\partial t}+v\frac{\partial v}{\partial x}=-\frac{\partial
\varphi }{\partial x}  \tag{1}
\end{equation}

\begin{equation}
\frac{\partial \sigma }{\partial t}+\frac{\partial }{\partial x}(\sigma v)=0
\tag{2}
\end{equation}

\begin{equation}
\frac{\partial ^{2}\varphi }{\partial x^{2}}=\sigma  \tag{3}
\end{equation}

where $v$ is the velocity, $\varphi $ is the gravitational potential and $%
\sigma $ is a quantity which is proportional to the product of the density
times the gravitational constant.$^{1)}$ Differentiating (1) with respect to 
$x$, with (3) taken into account, we obtain:

\begin{equation}
\frac{dh}{dt}+h^{2}=-\sigma ,\;\frac{d\sigma }{dt}+\sigma h=0,  \tag{4}
\end{equation}

\[
\frac{d}{dt}\equiv \frac{\partial }{\partial t}+v\frac{\partial }{\partial x}%
,\;h\equiv \frac{\partial v}{\partial x}. 
\]

The system (4) reduces to a single equation of the second order:

\begin{equation}
\frac{d^{2}}{dt^{2}}(\sigma ^{-1})=-1  \tag{5}
\end{equation}

In the Lagrangian description the solution has the form%
\begin{equation}
\sigma (t,x_{o})=\sigma _{o}(x_{o})[1+h_{o}(x_{o})-1/2\sigma
_{o}(x_{o})t^{2}]^{-1},  \tag{6}
\end{equation}

\begin{equation}
h(t,x_{o})=[h_{o}(x_{o})-\sigma _{o}(x_{o})t][1+h_{o}(x_{o})-1/2\sigma
_{o}(x_{o})t^{2}]^{-1},  \tag{7}
\end{equation}

where $x_{o}$ is the initial distance of a fluid particle from the fixed
particle to which the reference system is attached, $\sigma _{o}(x_{o})$ and 
$h_{o}(x_{o})$ denote the initial distributions.

We see that in a (1+1)-dimensional universe a contraction to a singularity
always occurs after a time

\[
t_{\ast }(x_{o})=\sigma
_{o}^{-1}(x_{o})\{h_{o}(x_{o})+[h_{o}^{2}(x_{o})+2\sigma
_{o}(x_{o})]^{1/2}\}, 
\]

although for $h_{o}>0$ at the beginning, of course, there will be an
expansion. One can show that the same result will also hold in a
(2+1)-dimensional universe.$^{2)}$

To formulas (6) and (7) it is necessary to add the time-dependence \ of the
moving coordinate $x(t,x_{o})$ of the fluid particle. It is easy to obtain
this dependence from the equation of continuity

\[
\frac{\partial x}{\partial x_{o}}=\frac{\sigma }{\sigma _{o}}, 
\]

which is another way of writing Eq. (2). We have

\begin{equation}
x(t,x_{o})=x_{o}+u_{o}(x_{o})t-1/2\mu _{o}(x_{o})t^{2},  \tag{8}
\end{equation}

\begin{equation}
u_{o}(x_{o})=\int\limits_{0}^{x_{o}}h_{o}(z)dz,\;\mu
_{o}(x_{o})=\int\limits_{0}^{x_{o}}\sigma _{o}(z)dz  \tag{9}
\end{equation}

Formulas (6), (7), (8), and (9) completely solve the problem in the
Lagrangian description. The same formulas in parametric form ($x_{o}$ is the
parameter) give the solution in the Eulerian description.

Let us emphasize the basic physical fact which enable us to obtain a simple
and exact solution: in a (1+1)-dimensional universe the relative
acceleration of the fluid particles is constant and is proportional to the
mass confined between the particles.

\textbf{3.} Let us consider the following example: a sinusoidal perturbation
of the initial density

\[
\sigma _{o}(x_{o})=<\sigma _{o}>(1+a_{o}\sin \frac{2\pi x_{o}}{L_{o}}%
),\;h_{o}=const, 
\]

where $<\sigma _{o}>$ denotes the average density, $a_{o}$ is the relative
initial amplitude, and $L_{o}$ is the initial spatial period. Formulas (6)
and (8) take the form

\[
\sigma (\tau ,s_{o})=<\sigma _{o}>(1+a\sin 2\pi s_{o})[1+\theta _{o}\tau
-1/2\tau ^{2}(1+a_{o}\sin 2\pi s_{o})]^{-1}, 
\]

\[
s\equiv \frac{x}{L_{o}}=s_{o}+s_{o}\theta _{o}\tau -\frac{1}{2}\tau
^{2}[s_{o}+\frac{1}{\pi }a_{o}\sin ^{2}\pi s_{o}], 
\]

\[
s_{o}=\frac{x_{o}}{L_{o}},\;\theta _{o}=h_{o}<\sigma _{o}>^{-1/2},\;\tau
=t<\sigma _{o}>^{1/2}. 
\]

The graphs are constructed in the Eulerian description for

\[
\theta _{o}=1,\;a_{o}=1/4,\;\tau =0,1,2. 
\]

The value $\tau =2$ corresponds to that moment when the spatial period,
after a temporary increase, is again equal to the initial period, but the
shape of the perturbation is now substantially distorted in accordance with
gravitational condensation.

The author wishes to thank Ya. B. Zel'dovich for a helpful discussion of
this work.

\bigskip

[1] L. D Landau and E. M. Lifshitz, Teoriya polya, Fizmatgiz, M. 1962 [The
Classical Theory of Fields, Pergamon Press, 1975].

[2] E. M. Lifshitz, Zh. Eksp. Teor. Fiz. \textbf{16}, 587 (1946).

[3] Ya. B. Zel'dovich and I. D. Novikov, Relyativistskaya astrofizika
(Relativistic Astrophysics), Nauka, M. 1967.

[4] Ya. B. Zel'dovich, Usp. Mat. Nauk \textbf{23}, 171 (1968).

\bigskip

--------------------------------

\bigskip

$^{1)}$ One can show that Eqs. (1) - (3) are the nonrelativistic limit of
Einstein's equations for a (1+1)-dimensional universe.

$^{2)}$ We present here one fantastic conjecture. Perhaps the universe was
not always (3+1)-dimensional. The dimensionality might change during a
transition through the singular state with zero space-dimensionality. Only
starting with space-dimensionality equal to three did the universe gain the
possibility "to survive".

\end{document}